\documentstyle[12pt,a4wide]{article}
\newcommand{\be}{\begin{equation}}
\newcommand{\ee}{\end{equation}}
\newcommand{\bea}{\begin{eqnarray}}
\newcommand{\eea}{\end{eqnarray}}
\begin{document}
\begin{flushright}
DO-TH-94/19\\
\end{flushright}

\vspace{20mm}

\begin{center}
{\Large \bf
What is the lightest excited state of the strongly selfcoupled
Higgs field?
}
\vspace{10mm}

{\large  V. G. Kiselev
\footnote{On leave of absence from Institute of Physics,
220602 Minsk, Byelorussia.

 ~~E-mail: kiselev@het.physik.uni-dortmund.de}
{and}
H. So
\footnote{On leave of absence from Dept. of Physics,
Niigata University, Niigata, Japan.

 ~~E-mail:  so@het.physik.uni-dortmund.de}
\footnote{The author to whom the correspondence should be addressed.}}
\\
\vspace{15mm}

{\large Institut f\"ur Physik, Universit\"at Dortmund} \\
{\large D - 44221 Dortmund, Germany}
\date{today}
\vspace{25mm}

{\bf Abstract}
\end{center}

We argue for the existence of an upper bound $m_*$
on the Higgs mass at which the lowest excited state of the Higgs
field ceases to be the conventional plane wave.
An explicit construction of an alternative nonperturbative state
is discussed.
This excitation is spatially localized. The field fluctuations inside
the localization region are large.
The energy of the excitation is smaller than the mass of the
plane wave state $m$ at $m>m_*$.
An approximate value of $m_*$ is found to be
$m_* \approx 4.75$ times the vacuum expectation value.
This is an upper bound at the tree level.

\setcounter{footnote}{0}
\newpage
Let us consider the one-component Higgs field
specified by the action
\be                              \label{S}
S=\int d^4x \left[
{1\over2}(\partial_\mu \varphi)^2 - U(\varphi)
\right]
\ee
where the potential term, $U(\varphi)$, is defined as
\be
U(\varphi) = {\lambda}(\varphi^2-v^2)^2.
\ee
The canonical quantization suggests that the minimal-energy excitation
above the vacuum is the one-particle state with the mass squared
$m^2=8 \lambda v^2$. This conclusion is based on the small fluctuation
analysis on the fixed classical background $\varphi=v$. Any feedback
effect is neglected. This is valid for small values of the coupling
constant because the energy associated with deviations of the
classical $\varphi$-field
from its vacuum value is typically $m/\lambda \gg m$.
But there is no reason to expect the Higgs particle with
the mass about $\sqrt{8\lambda}v$ at large values of $\lambda$.
Moreover, some indications of nontrivial excitations in the Higgs
sector have been collected \cite{Do}.

In this paper we discuss a scenario of the departure from the canonical
picture at large $\lambda$. We present some steps to the explicit
construction of a nonperturbative excited state. As it shown by
the estimates presented below this state is lighter than the
mass of the conventional plane wave at $m>m_* \approx 4.75v$.
Unlike the plane wave, it is localized in a region of the order of
(0.1 -- 0.3) $v^{-1}$. The mean value of the Higgs field inside
this volume is smaller than $v$.
The field fluctuations there are of the order of $v$.

A very naive approach to treat such a state is to take some space
dependent field configuration $\varphi=\Phi(x)<v$ with
$\Phi(\infty)=v$ and
consider the small-fluctuation spectrum around it.
There is, however, a well-known fact that any classical configuration
 collapses to the origin in the usual Higgs system \cite{col}.
 To overcome this, we shall consider   the
lowest eigenvalue, $\Omega_0^2$, of this spectrum.
Then one can expect to find
the energy of the lightest excited state ${\cal E}$ as the minimum of
$E[\Phi(x)]+\Omega_0[\Phi(x)]$.
Such a minimum can only be reached when $\Omega_0 < m$ corresponds
 to a bound state in the potential created by $\Phi(x)$.
While the order of magnitude of $\Omega_0$ is $m$, the value of $E$
 is typically  $m/\lambda$. These two values can balance each
other at large $\lambda$ only.

A crucial question that one has to address to this plan is how to make
distinguish between the background field and the fluctuations around
it. Moreover, it is clear that one has to treat both of them
simultaneously performing the quantization if the coupling constant
is large.
An approach presented here allows us to proceed toward this aim.

Let us start with an illustrative  model. Consider a system with two
degrees of freedom described by the Schr\"odinger equation
\be                                 \label{model}
\left(
- {1\over 2\mu}{\partial^2\over \partial x^2}
- {1\over 2}{\partial^2\over \partial y^2}
+ V(x) + {1\over 2}\omega^2(x) y^2
\right)\Psi(x,y) = {\cal E} \Psi(x,y) .
\ee
Suppose $V(x)$, $x\ge 0$, is a monotonously increasing function,
$V(0)=0$.  Unlikely, the function $\omega^2(x)$ decreases.
So, the full potential in eq. (\ref{model}) looks like a valley that
leads upward, but
becomes wider as $x$ increases.  What are the stationary states of
this system?  Do the wave functions concentrate near the origin?

It is very easy to answer these questions in the adiabatic
approximation that is valid if the motion in the $x$ direction is much
slower than that in $y$. In this case the wave function can be found
in the form
\be                                 \label{Psimod}
\Psi(x,y) = f_n(y;\omega(x))\psi(x)
\ee
where $f_n(y;\omega)$ is the $n$-th normalized eigenfunction of the
harmonic oscillator with the frequency $\omega$. One substitutes
(\ref{Psimod}) to the equation (\ref{model}), then multiplies
it by $f_n$ and integrates over $y$ taking into account that
$\int d y f_n(y){\partial f_n(y)/\partial \omega}=0$. It yields an
effective Schr\"odinger equation for $\psi(x)$:
\be                                 \label{model'}
\left(
- {1\over 2\mu}{\partial^2\over \partial x^2}
+ V(x) + (n+{1\over 2})\omega(x)
\right)\psi(x) = {\cal E} \psi(x) .
\ee
One observes that an effective potential $V_{eff}=V+(n+1/2)\omega $
enters this equation. The minimum of $V_{eff}$ may be far from the
point $x=0$.
In other words, the wave function is repulsed from the region
where the potential is too narrow.

What is the first correction to the adiabatic approximation?
In order to find it out one substitutes
$\Psi(x,y)=\psi(x)f_n(y;\omega(x))$
into the equation (\ref{model}) and finds the first order
correction $V^{(1)}$ to $V$.
It reads
\be                            \label{adia}
V^{(1)}
= -  {1\over 2\mu } \left( \frac{d\omega}{dx} \right)^2
\int dy
f_n(y;\omega(x))
\frac{\partial^2 f_n(y;\omega(x))}{\partial \omega^2}.
\ee
If one takes into account that ${\cal E} \sim n\omega $ then
the correction ($\ref{adia}$) divided by ${\cal E}$  turns into
$V^{(1)} / {\cal E} \sim n/(\omega \mu \Delta x^2)$.
Here $\Delta x$ denotes the characteristic length at which $\omega$
changes essentially. $\mu \Delta x^2$ is the typical time of the
motion over this interval.

Let us discuss the relation of this model to our problem.
We would like to construct a valley in the field configuration
space similar to that formed by the potential in eq. (\ref{model}).
A coordinate $X$, an analog of $x$, parametrizes the bottom of
this valley. The point $X=0$ corresponds to $\Phi=v$. The larger $X$
the farther $\Phi(x)$ deviates from the vacuum expectation value.
The curvature of the valley cross section is to be smaller as $X$
increases. As we see soon it is related to $\Omega_0$,
the lowest normal frequency of the fluctuations around $\Phi(x)$.
This suggests a choice of the coordinate $X$. We take just
\be                 \label{X}
X = m - \Omega_0.
\ee

With this choice we can specify the bottom of the valley we are going
to consider. It is the configuration $\Phi(x)$ at which the classical
field energy
\be                       \label{E}
E[\varphi(x)] =
\int d^3x \left[
{1\over2}(\vec{\nabla}\varphi)^2 + U(\varphi)
\right]
\ee
is minimal at fixed value of $\Omega_0$. Let us write this condition
in the form
\be                               \label{min}
E[\Phi(x)] = \min_{{\rm fixed~ \Omega_0}} E[\varphi(x)].
\ee
Here $\Omega_0$ is the lowest eigenvalue of the problem
\be                           \label{SE}
\left(
- \Delta + m^2 + U''(\varphi)
\right)\psi(x) = \Omega^2 \psi(x).
\ee
The corresponding eigenfunction $\psi_0(x)$ is spherically
symmetric and nodeless.

Let us specify an analog of the coordinate $y$ of our quantum
mechanical example (\ref{model}). It is convenient to this end to
expand the field near the bottom of the valley
\be
\varphi(x) = \Phi(x) + \chi(x)
\ee
and consider the first and the second variations of the classical
energy, $\delta E$ and $\delta^2 E$ up to the terms of the order
$\chi^2$.

All the functions $\chi(x)$ constitute a real linear space,
${\cal H}$, with the inner product
\be
(\chi_1,\chi_2)\equiv\int d^3x \chi_1(x) \chi_2(x).
\ee
There are two specific coordinate systems in it. The introduction of
the first one relates to $\delta^2 E$. This value reads
\be
\delta^2 E = \int d^3x \chi(x) A(\Phi(x)) \chi(x).
\ee
Here is $A(\Phi(x))$ the operator in the LHS of eq. (\ref{SE}).
The eigenfunctions of $A$ form a complete set. Let us refer to these
functions, the corresponding eigenvalue and coordinates as,
respectively, $\psi_\alpha (x)$, $\Omega_\alpha^2$, and $\xi_\alpha$,
$\alpha=0,1,\dots$.

The other set of coordinates includes the direction of the
bottom of the valley defined by (\ref{min}) as one of the basis
vectors.
This set can be chosen as the eigenfunctions of an operator
$\tilde{A}=P A P$ where $P$ is the projector on $\Sigma$,
the hypersurface of fixed $\Omega_0$ at $\varphi(x) = \Phi(x)$.
The vector that is tangential to the bottom of the valley
in the configuration space is just the gradient of the energy
$\psi_E(x)$.
It is orthogonal to $\Sigma$ due to the condition (\ref{min}) and,
therefore, it is an eigenfunction of $\tilde{A}$ with zero eigenvalue.
Its explicit form is
\be                       \label{grad}
\psi_E(x) =
-\Delta \Phi(x) + U'(\varphi).
\ee
In terms of this function the first differential of the energy
for the direction $\chi(x)$ in ${\cal H}$
reads
$\delta E$,
\be
\delta E = \int d^3x \psi_E(x) \chi(x).
\ee
All other eigenfunctions of $\tilde{A}$
denoted by $\chi_\beta$, $\beta=0,1,\dots$, belong to $\Sigma$.
Let the corresponding eigenvalues and the coordinates be
$\omega_\beta$ and $\eta_\beta$.
Note that for each $\chi(x)$ in $\Sigma$ the mean value
$(\chi,\tilde{A}\chi) = (\chi,A\chi)$. The minimal
$(\chi,A\chi)$ among all normalized $\chi$ in $\Sigma$ is just
$\omega_0^2$, the minimal nonzero eigenvalue of $\tilde{A}$.

Let us go on constructing the valley similar to one in (\ref{model}).
To this end we make use of a trial wave functional
\be                      \label{Psi}
\Psi[\eta_\beta] = \psi (X) f_1(\eta_0;\omega_0)
\prod_{\beta=1}^{\infty} f_0(\eta_\beta;\omega_\beta).
\ee
The average energy of the quantum motion in the hypersurface
$\Sigma$ given by (\ref{Psi}) is
\be                      \label{sums}
{3\over2} \omega_0
+ \sum_{\beta=1}^{\infty} {1\over 2} \omega_\beta
\ee
minus the vacuum zero energy.
If we add to this sum one more frequency related to the
motion along the valley then the expression (\ref{sums}) turns to
$\omega_0$ plus the one-loop correction to $E[\Phi(x)]$.
A  preliminary result for the one-loop corrections is discussed later
 and we neglect them now.
To this accuracy, the energy of the excited Higgs field takes the form
\be                                  \label{Eeff}
E_{eff}[\Phi(x)] = E[\Phi(x)]  + \omega_0(\Omega).
\ee
In the following we refer to this value as the bottom of the effective
valley.

What is the value of $\omega_0$?
If we are lucky the gradient $\psi_E$ is orthogonal to $\psi_0$.
In this case we reach the minimal value $E_{eff}=E+\Omega_0$.
In the opposite limiting case of the bad luck $\psi_E \sim \psi_0$.
Then the minimal  value of the effective energy is
shifted to $E_{eff}=E+\Omega_1$.
This implies  the nontrivial behaviour of the Higgs configuration
does not appear till larger values of the Higgs mass where our
possible approximations are not good.

The intermediate case can be illustrated in a simplified way if we
consider in addition to $\xi_0$ only one coordinate $\xi_1$
corresponding
to the eigenvalue $\Omega_1^2$. Let the projections of $\psi_E$
on $\psi_0$ and $\psi_1$ be proportional to $\cos \theta_{0E}$ and
$\sin \theta_{0E}$. Then
\be                                \label{corr2}
\omega_0^2 = \Omega_0^2 \sin^2 \theta_{0E}
+ \Omega_1^2 \cos^2 \theta_{0E}.
\ee
One sees that a small value of $\cos \theta_{0E}$ yields a quadratic
correction to $\omega_0$:
\be                               \label{corr}
\omega_0 \approx
\Omega_0 + \frac{\Omega_1^2-\Omega_0^2}{2 \Omega_0} \cos ^2\theta_{0E}.
\ee
In reality we have to consider the complete set of coordinates and
eigenvalues together with $\Omega_1$. Those eigenvalues are typically
of the order $m$. The expression (\ref{corr}) is, of course, only
an order-of-magnitude estimate.

Let us present our results.
In order to probe the structure of $E$ in the configuration space we
use the following spherically symmetric ansatz for the $\varphi$-field
(fig.1):
\be                                \label{Ansatz}
\varphi(r) = v \left[
1 - b \left( 1+\frac{r^2}{a^2} \right) ^3
\exp \left( - \frac{r^2}{a^2} \right)
\right].
\ee
This trial function has two parameters $a$ and $b$
related to the width and the amplitude of the configuration.

It is straightforward to find the bottom of the valley (\ref{min}).
We do it by the variation of $a$ and $b$ and searching for the minima
in small $\Omega_0$ intervals.
Typically the total number of data point is a few ten-thousands.
A few tens from them are chosen.
Some results of this procedure are presented in figs. 2 -- 5.
The valley shapes in the $(a,b)$-plane  are shown in
fig. 6.

The values of the cosine of the angle between $\psi_0$ and
$\psi_E$
\be                  \label{cos}
\cos \theta_{0E} = \frac{(\psi_0, \psi_E)}{\sqrt{(\psi_E,\psi_E)}}
\ee
are also shown in figs. 2 -- 5. It turns out that the ansatz
(\ref{Ansatz}) is sufficiently good in the sense that it yields
small values of
$\cos \theta_{0E}$ except for the region of $\Omega_0$
close to zero. In order to estimate the correction
to $E_{eff}$ due to nonorthogonality of $\psi_E$ and $\psi_0$
we make use of the equation (\ref{corr2})
with $\Omega_1=m$. The resulted increase of $E_{eff}$ near
$\Omega_0 \approx 0$ is seen in figs. 2 -- 5.
The value of $\cos \theta_{0E}$ up to 0.3 can be neglected with,
approximately, one per cent accuracy in $\omega_0$ (\ref{corr}).

The region $\Omega_0 \approx m$ can not be described at all with our
definition of the valley (\ref{min}).
The reason is that the value of $\Omega_0$ remains
$\Omega_0=m$ as long as the potential in eq. (\ref{SE}) is too weak to
give rise to a bound state. In particular it means that $\Omega_0$
is a nonanalytic function of $a$ and $b$ as they are small.
Note that the potential in (\ref{SE}) can be treated perturbatively
in this case.

Let us discuss the results.
If the Higgs particle is light, then $E_{eff}$ monotonously increases
 as $\Omega_0$ gets smaller. This confirms that the  lightest
 excited state is the conventional plane wave.
 A nontrivial local minimum of $E_{eff}$ first appears at
 $m \approx 4.38v$.
 It is illustrated with fig. 2.

The bottom of the effective valley (\ref{Eeff})
becomes degenerate with the plane wave energy $m$ at $m=4.75v$
(fig. 3). One can expect large fluctuations in the $\Omega_0$
direction (we discuss
the width of the appropriate wave function below).
The lowest excited state ceases to be the plane wave.

A minimum of $E_{eff}$ appears and becomes deeper as $m$
increases (figs. 4, 5).
In $m>m_*$ we use the parameter $m$ just instead of less manifest
value of $\lambda=m^2/8v^2$. The actual energy of the lowest excited
state ${\cal E}$ is less than $m$. Its value without any corrections
due to the quantization of $\Omega_0$ is just $E_{eff}$ at the
minimum.
It is plotted in fig. 7 as a function of $m$.

The value of $E_{eff}$ at the left edge of the effective valley, i.e.
at $\Omega_0\approx 0$, becomes smaller than at the local minimum
$\Omega_0^{min}$ at $m\approx 6.1 v$ (fig. 7).
At $m\approx 6.78v$ the minimum of $E_{eff}$ at $\Omega_0^{min}$
merges with a local maximum and disappears. The bottom of
the effective valley at larger $m$ decreases monotonously
as $\Omega_0$ gets smaller. This could be a hint on a the more
complicated
configuration with even lower energy. Unfortunately, our data do not
allow to make such a conclusion. The correction to $E_{eff}$ due to
the nonorthogonality of $\psi_0$ to $\psi_E$ shown in figs. 2 -- 5
increases $E_{eff}$ near $\Omega_0\approx 0$. This can be either a
meaningful behaviour or an artifact of our ansatz.

The quantization of the motion along the valley is not
straightforward.
Doing it properly we have to consider a nonlocal action that arises
after all the degrees of freedom across the valley are integrated out.
It is the adiabatic approach that allows us to neglect this
complication.
Thus, a naive way is just to make use of the Schr\"odinger equation
\be                                 \label{EQ}
\left(
- {1\over 2\mu_\Omega}{\partial^2\over \partial \Omega_0^2}
+ E_{eff}[\Phi(x)]
\right)\psi(\Omega_0) = {\cal E} \psi(\Omega_0)
\ee
with
\be                              \label{mu}
\mu_\Omega (\Omega_0) = \int d^3 x
\left(
\frac{\partial \Phi}{\partial a}\frac{d a}{d \Omega_0}
+ \frac{\partial \Phi}{\partial b}\frac{d b}{d \Omega_0}
\right)^2.
\ee
We face, however, the problem of ordering because this mass parameter
$\mu_\Omega$ depends on the coordinate, $\Omega_0$.
In our case, we first apply the differential operator to the wave
function then multiply it by $1/\mu_\Omega$. This naive choice gives
the qualitative character to our discussion.

In spite of the poor accuracy of the equation (\ref{EQ}) we use
it in order to make some estimates. First of all it is worth
to check  the validity of the adiabatic approximation.
Its parameter can be chosen as $\delta_1=\nu / \Omega_0$ where
$\nu = (E_{eff}''/\mu_\Omega)^{1/2}$ is the frequency of the small
oscillations around the minimum of $E_{eff}$. Some values of
$\delta_1$ are collected in table 1.
We also present there $\delta_2$, the maximum -- by absolute value --
of the ratio of $V^{(1)}$ eq. (\ref{adia}) to $E_{eff}$,  with
$x=m-\Omega_0$, $\omega=\Omega_0$ and $n=1$.
As it is discussed below the minimum of $E_{eff}$ can hardly be
considered as a harmonic oscillator. This suggests that
the parameter $\delta_2$ is more relevant than $\delta_1$.
The smallness of $\delta_2$ and not too large value of $\delta_1$
let us hope that our results are qualitatively valid. We have to
remind, however, that the values of both parameters depend on the form
of the kinetic term in the equation (\ref{EQ}).

The further use of this equation allows us to estimate the
characteristic
distance and time of the motion in the effective valley.
Let $\Delta \Omega_0$ be the fluctuation of $\Omega_0$.
One sees from fig. 3 that the bottom of the effective valley
(\ref{Eeff}) is flat at $m=4.75v$. So, one can expect
the large fluctuations $\Delta \Omega_0\approx m$.
As the Higgs mass is larger than $4.75v$, $E_{eff}$
looks like an oscillator potential rather than a flat
one (figs. 4, 5).
Is it a harmonic oscillator? If it is the case
then the wave function of the motion along the valley is
$\psi(\Omega_0)\approx f_0(\Omega_0;\nu)$ and
$\Delta \Omega_0 \approx ({2/(\mu_\Omega \nu)})^{1/2}$. The numerical
results collected in table 1 show that it is not the case, especially
at $m\approx m_*$.
The boundary condition for $\psi(\Omega_0)$ at
the edges of the effective valley should be also essential.

A characteristic time $\tau_{free}$ for the flat $E_{eff}$ can be
estimated by the use of the Green function of the
one-dimensional free motion. If one creates a well localized state
near $\Omega_0=m$ then the time of the wave function propagation on
the characteristic distance $\Delta \Omega_0$ is $\tau_{free}\approx
\mu_\Omega \Delta \Omega_0^2/2$. Its numerical values are
given in table 1 for $\Delta \Omega_0 = \Omega_0^{min}$.
One observes that this time is shorter than the inverse of
the excitation energy only at $m=4.38v$. At larger mass a possible
propagation from the plane wave state to $\Omega_0^{min}$ takes the
time which is longer than $1/{\cal E}$.
Another estimate of $\tau$ can be done for $m>m_*$ in the form
$\tau_{osc}\approx 2\pi/\nu$. This is comparable with the values of
$\tau_{free}$ at $m=5.5v$ and $m=6.78v$ (table 1).

To sum up, we have shown by an explicit construction at the tree level
that there is an upper bound $m_*$
on the Higgs mass at which the lightest excited state ceases to be the
conventional plane wave. This state becomes to be spatially localized
 and has the mass smaller than the naively expected Higgs mass $m$
(fig. 7). We have found an approximate upper bound of
$m_* \approx 4.75v$. In the minimal standard model it would correspond
to $m \approx 1.17$ TeV.

Let us discuss briefly some corrections to our results.
The effective valley (\ref{Eeff})
appears in balancing  the energy $E$ of the classical
field with the lowest excitation $\Omega_0$ on its background. This,
obviously, violates the basis of the usual loop expansion. Normally
such corrections should be small. Another approximation we have used
is the adiabatic approach. It does not require the effect to be small,
but needs the motion along the valley to be much slower than that
across.
In the table 2 we present our preliminary estimates of the one-loop
corrections to (\ref{E}) obtained by the method described in
\cite{BK}. These data indicate relatively good quality of the tree
approximation.
Note that both the one-loop correction to $E$ and the
correction (\ref{adia}) are negative.
So, they make the valley deeper\footnote{
We neglect here the one-loop contributions from other particle
species.
The Higgs particle in our consideration is much heavier than the
top-quark, the heaviest particle of the standard model \cite{top}.
Its contribution to $E[\Phi(x)]$ can be found by the method of ref.
\cite{bss}.
 }.
We have neglected also the anharmonicity of the fluctuations around
$\Phi(x)$. This obviously corresponds to the omission of the
loop contributions to $\Omega_0$.

Let us note that the introduction of the valley in the functional
space is a step to  distinguish between the classical
background and the fluctuations around it.
Indeed, it allows us to perform simultaneously
the quantization of the coordinate related to the background
and the fluctuations if one leaves behind the adiabatic approach.

Another source of correction is our use of the ansatz. Normally,
when one proceeds in this way, the resulted value of the energy
is an upper bound on the true minimum.
An ansatz-free calculation of the bottom-of-the-valley energy will be
published elsewhere.

Are our results applicable to the Standard Model of the electroweak
interaction? The answer based on the data presented here is rather
"no" because of too large threshold value of the Higgs mass
$m_* \approx 4.75v \approx 1170$ GeV. This is near the number at which
the Higgs decay width becomes equal to its mass \cite{Higgs}. However,
the  sources of the corrections discussed above give the
negative contributions to the excitation energy. If they result in
essential decreasing of $m_*$ then the consideration of the nontrivial
Higgs field excitation in the Standard Model can become meaningful.

It would be interesting to relate the phenomenon we have discussed
with the so-called triviality of the Higgs field \cite{triv}.
This question is, however, outside the scope of this paper.

We are grateful to Alexander von Humboldt Foundation for the support
of this work. V. G. K. thanks K. Kang for a discussion.

\newpage

\noindent{\Large \bf Tables }

\begin{table}[h]
\begin{center}
\begin{tabular}{|c|c|c|c|c|c|c|c|}
\hline
$\displaystyle{m/v} $&
$\displaystyle{{\cal E}/v}$ &
$\displaystyle{\Omega_0^{min}/ v}$ &
$\displaystyle{\Delta \Omega_0 / v}$ &
$\displaystyle{(\tau_{free}v)^{-1}}$ &
$\displaystyle{(\tau_{osc}v)^{-1}}$ &
$\displaystyle{\delta_1}$ &
$\displaystyle{\delta_2}$
\\
\hline
4.38 & 4.56 & 4.02 & 2.86 & 48 & $2.9 \times 10^{-1}$ &  0.40
&  0.044      \\
4.75 & 4.75 & 3.30 & 2.77 & 2.6     & $1.4 \times 10^{-1}$
&  0.27 & 0.057      \\
5.50 & 4.98 & 3.12 & 1.40 & $2.9 \times 10^{-1}$
& $9.1 \times 10^{-2}$    &  0.19 &  0.050      \\
6.78 & 5.09 & 2.42 & 1.82 & $6.7 \times 10^{-2}$
& $5.9 \times 10^{-2}$   &  0.15 &  0.144      \\
\hline
\end{tabular}
\end{center}
\caption{Some parameters of the effective valley defined in the text.
$\Omega_0^{min}$ means the value of $\Omega_0$ at the minimum of
$E_{eff}$.}
\end{table}

\bigskip

\begin{table}[h]
\begin{center}
\begin{tabular}{|c|c|c|c|c|c|}
\hline
$\displaystyle{m/v} $&
$\displaystyle{av}$ &
$\displaystyle{b}$ &
$\displaystyle{{\cal E}/v}$ &
$\displaystyle{{\cal E}_{1-loop}/v}$&
$\displaystyle{{\cal E}_{1-loop}/{\cal E}}$
\\
\hline
4.38 & 0.183 & 0.100 & 4.56 & -0.058 &  -0.013 \\
4.75 & 0.163 & 0.216 & 4.75 & -0.329 &  -0.069 \\
5.50 & 0.149 & 0.271 & 4.97 & -0.691 &  -0.139 \\
6.78 & 0.144 & 0.305 & 5.09 & -1.854 &  -0.363 \\
\hline
\end{tabular}
\end{center}
\caption{
The values of $a$, $b$, and the one-loop correction to ${\cal E}$
at the minimum of $E_{eff}$ at different $m$. The values of ${\cal E}$
are presented again for the reader's convenience.
}
\end{table}

\newpage

\noindent{\Large \bf Figure Captions}

\begin{description}
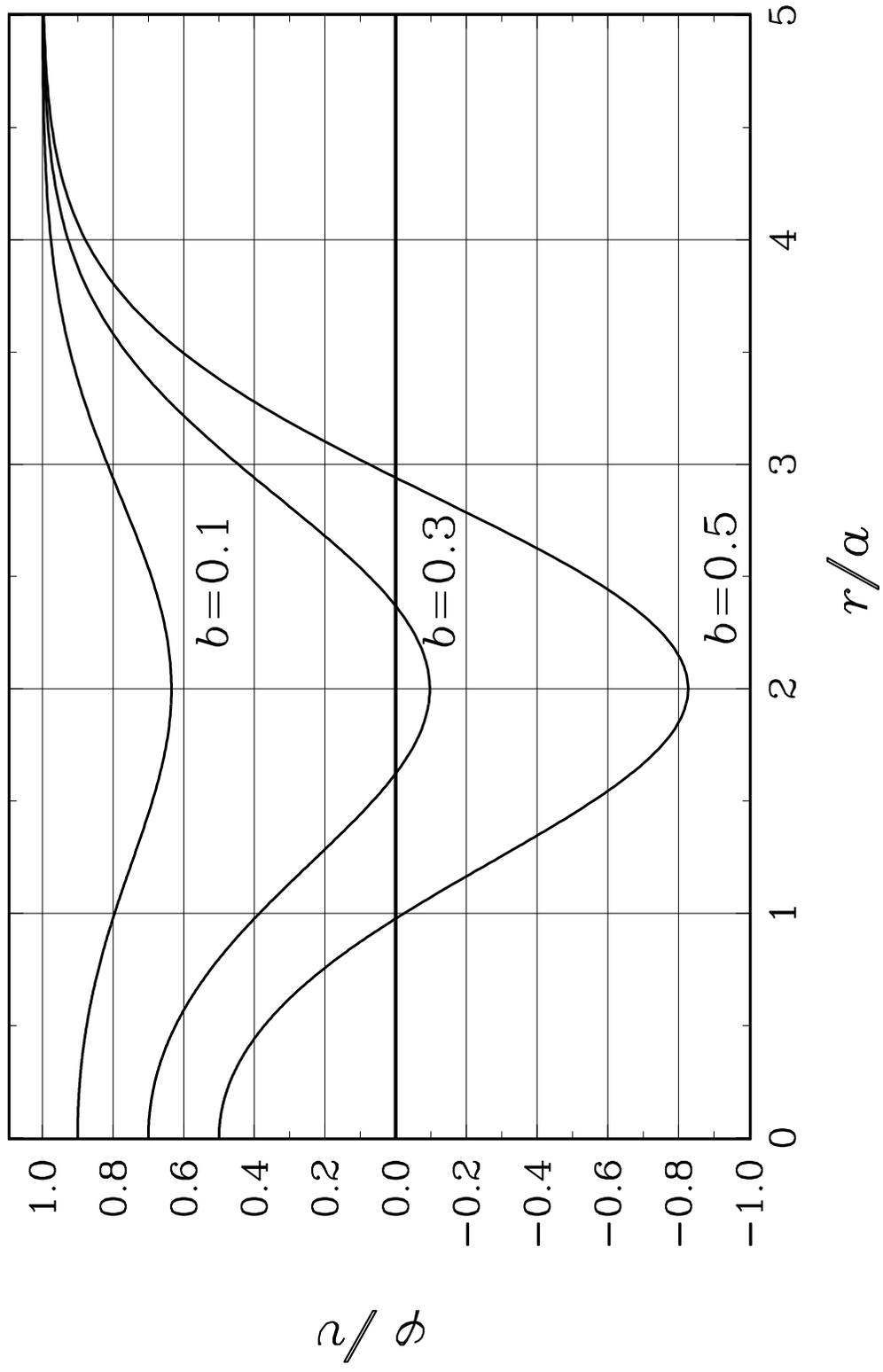
\item[Fig. 1] The shape of $\phi(x)$ given by the ansatz
(\ref{Ansatz}) at different values of $b$.
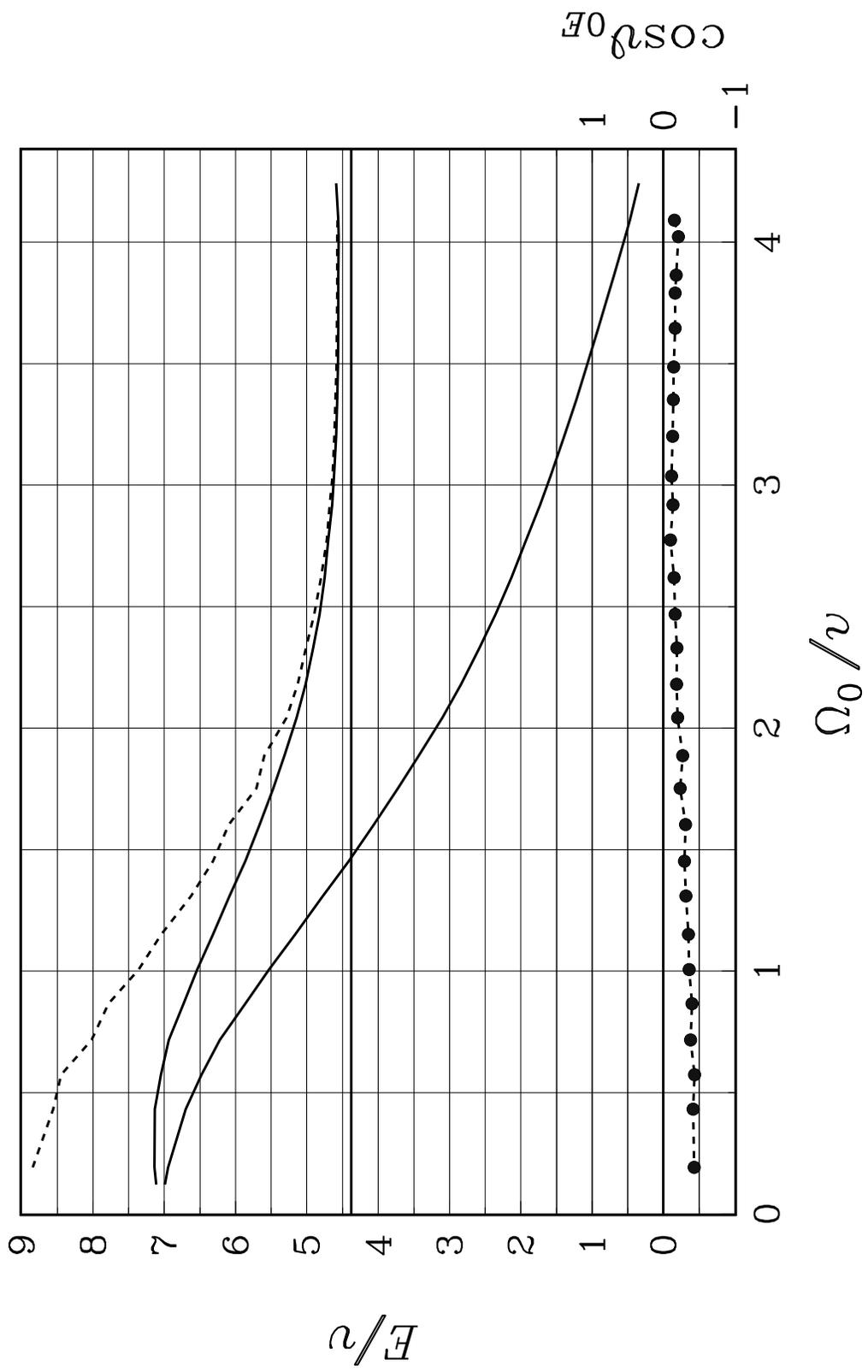
\item[Fig. 2] The bottom of the valley $E[\Phi(x)]$ defined in
eq. (\ref{min}) (the lower solid line), the bottom of the effective
valley (\ref{Eeff}) with $\omega_0=\Omega_0$
(the upper solid line), the cosine of the angle between $\psi_E$ and
$\psi_0$ (\ref{cos}) (short-dashed line), and the value
of $E_{eff}$ corrected accordingly to eq. (\ref{corr2})
(long-dashed line) at $m=4.38v$.
The data points are shown in the short-dashed line.
The thick horizontal line indicates the energy equal $m$.
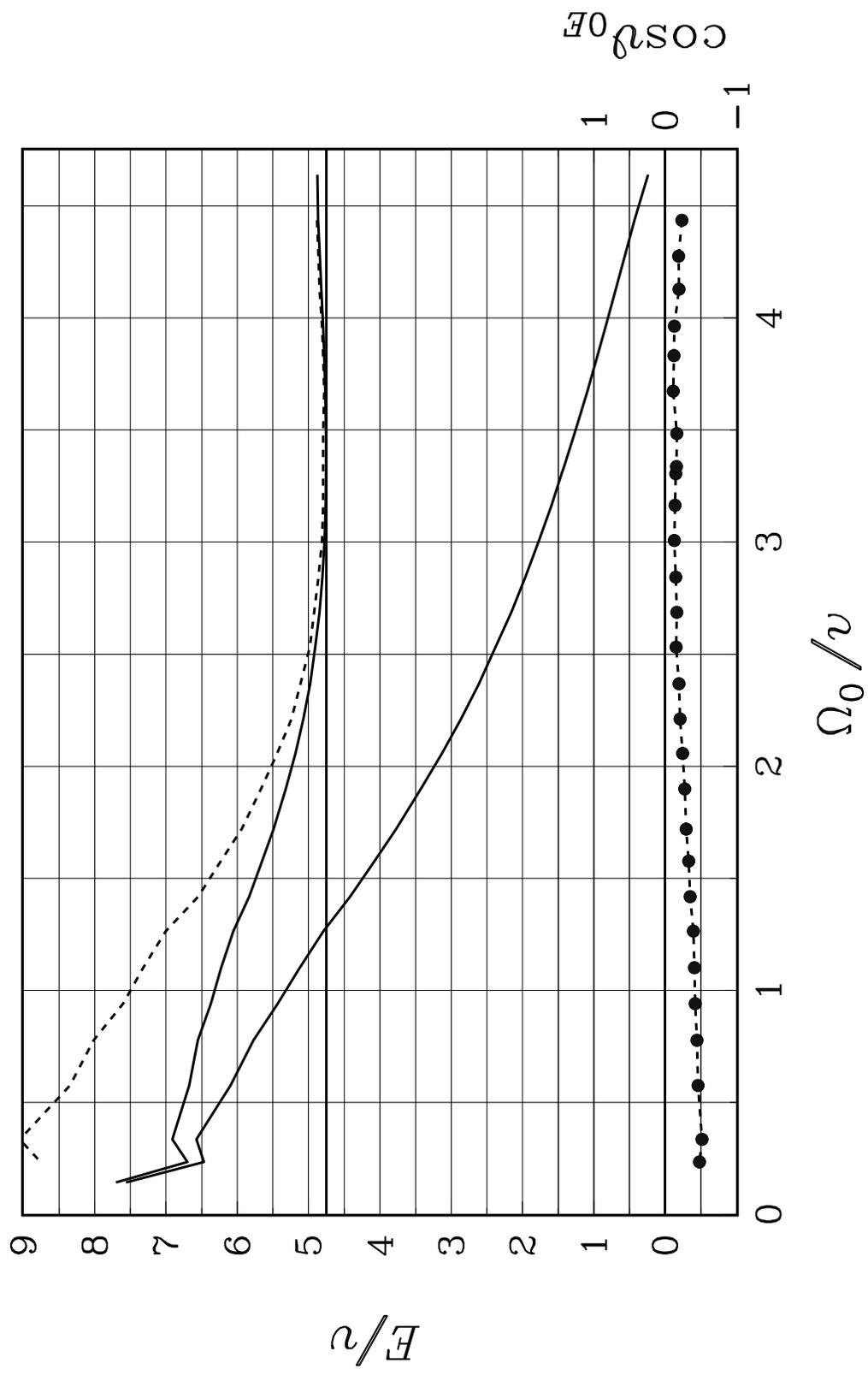
\item[Fig. 3] The same as in fig. 2 at $m=4.75v$.
The irregular behaviour at small $\Omega_0$ is due to the poor 
statistics collected in this region.
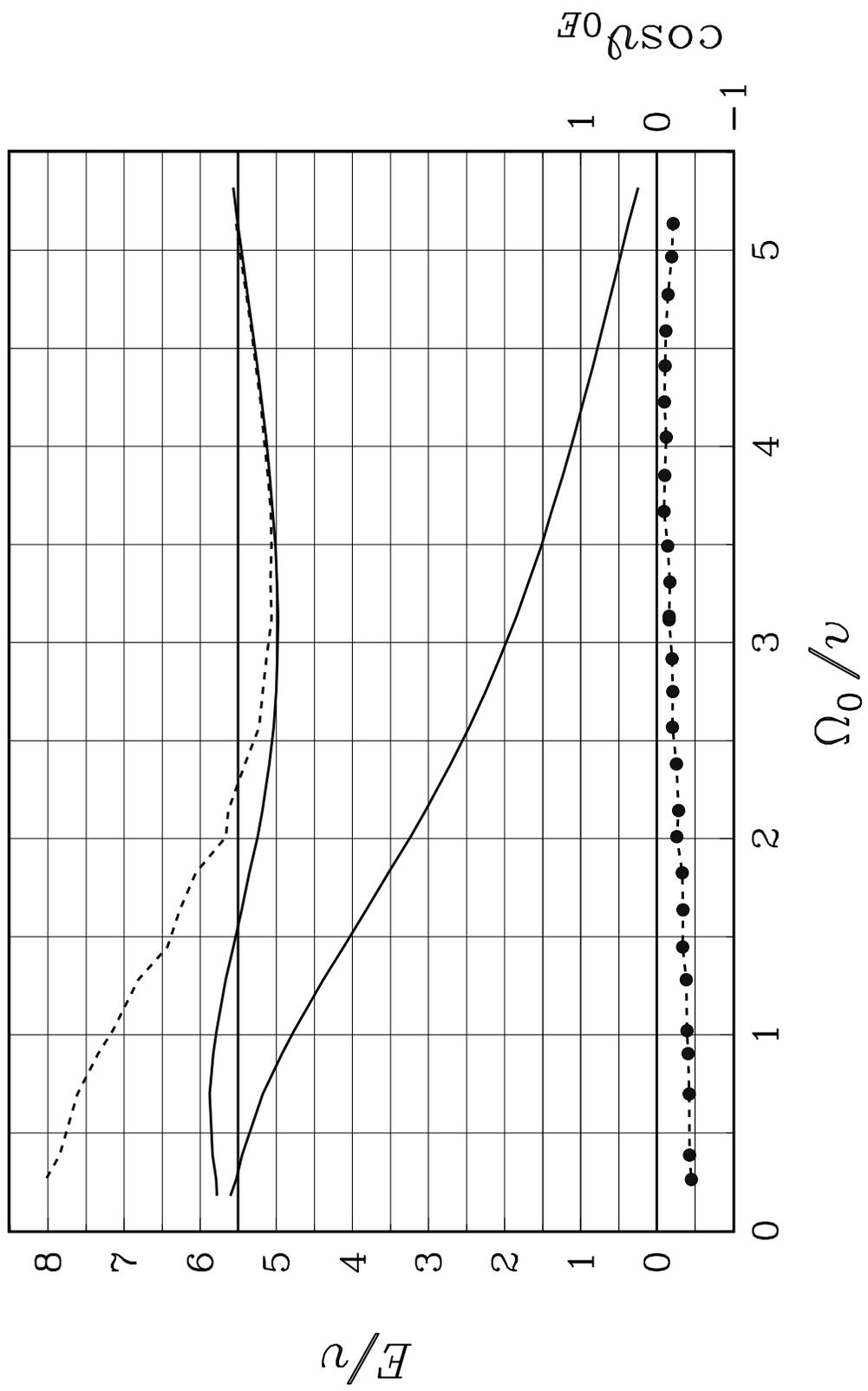
\item[Fig. 4] The same as in fig. 2 at $m=5.5v$.
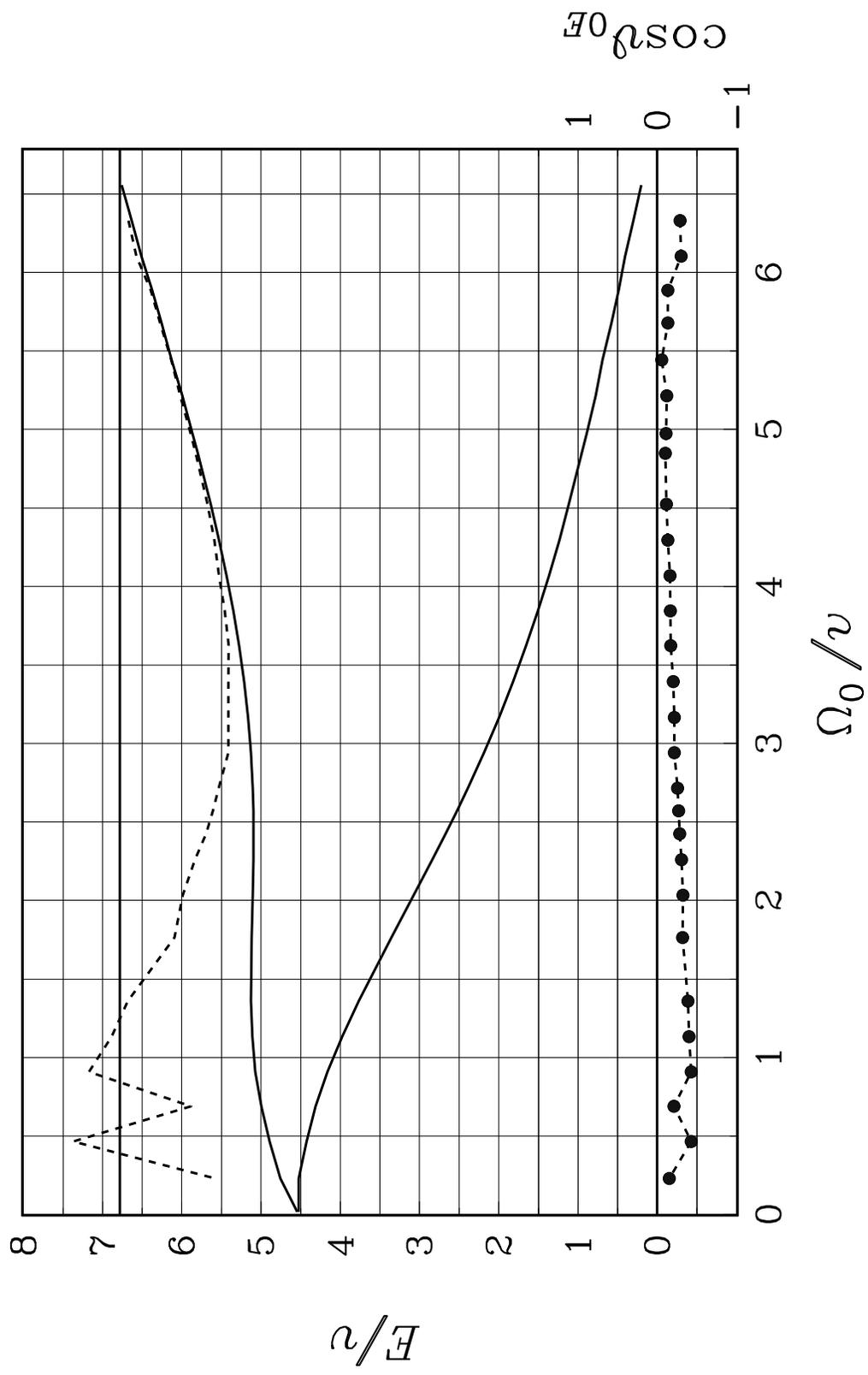
\item[Fig. 5] The same as in fig. 2 at $m=6.78v$. See also the 
caption to fig. 3.
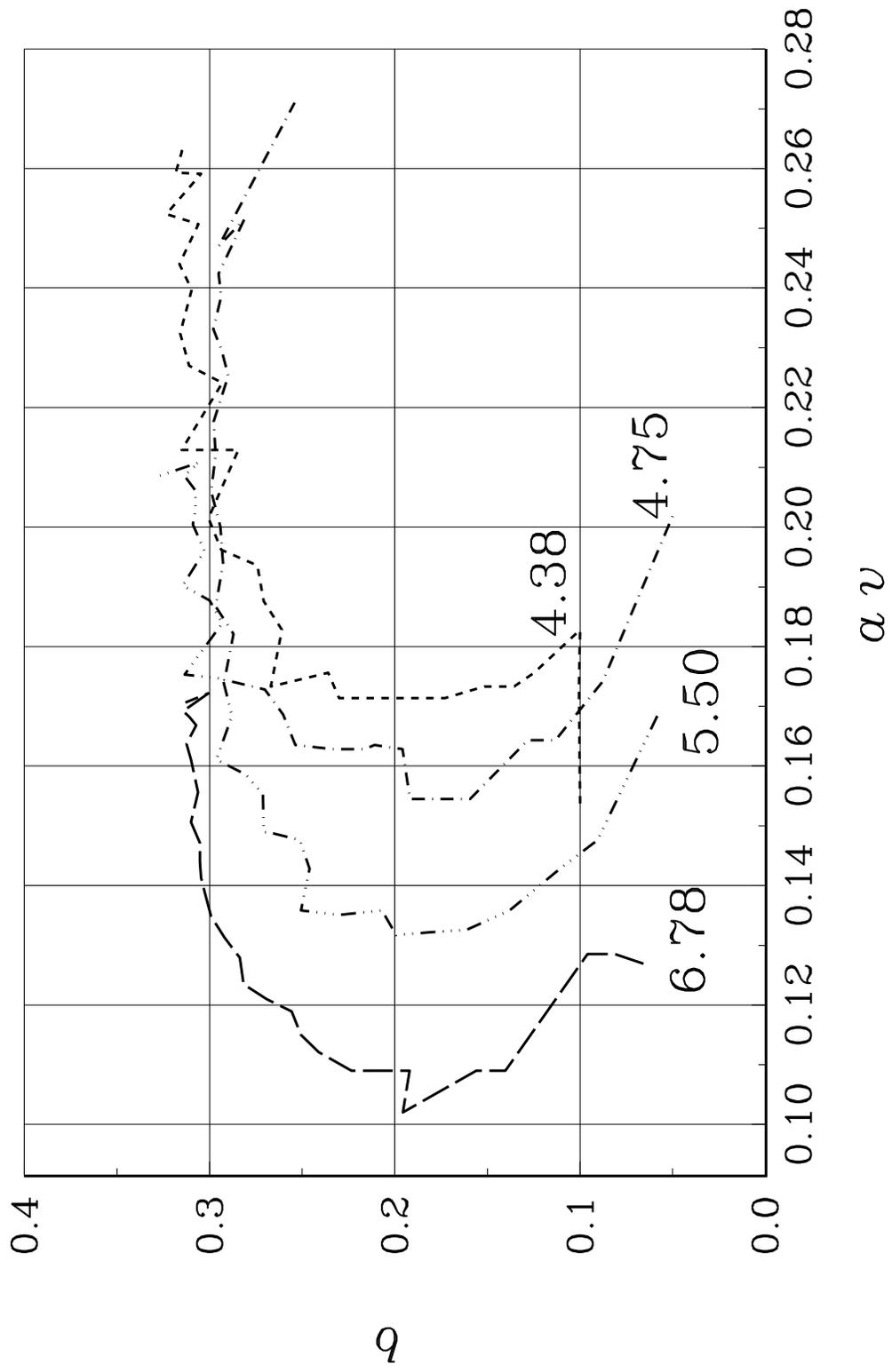
\item[Fig. 6] The valley shapes in the $(a,b)$-plane. Four curves
are labelled with the value of $m/v$. 
Note that the relatively large error in $a$ and $b$ yields 
an essentially smaller variation of $E_{eff}$ computed near its 
minimum.
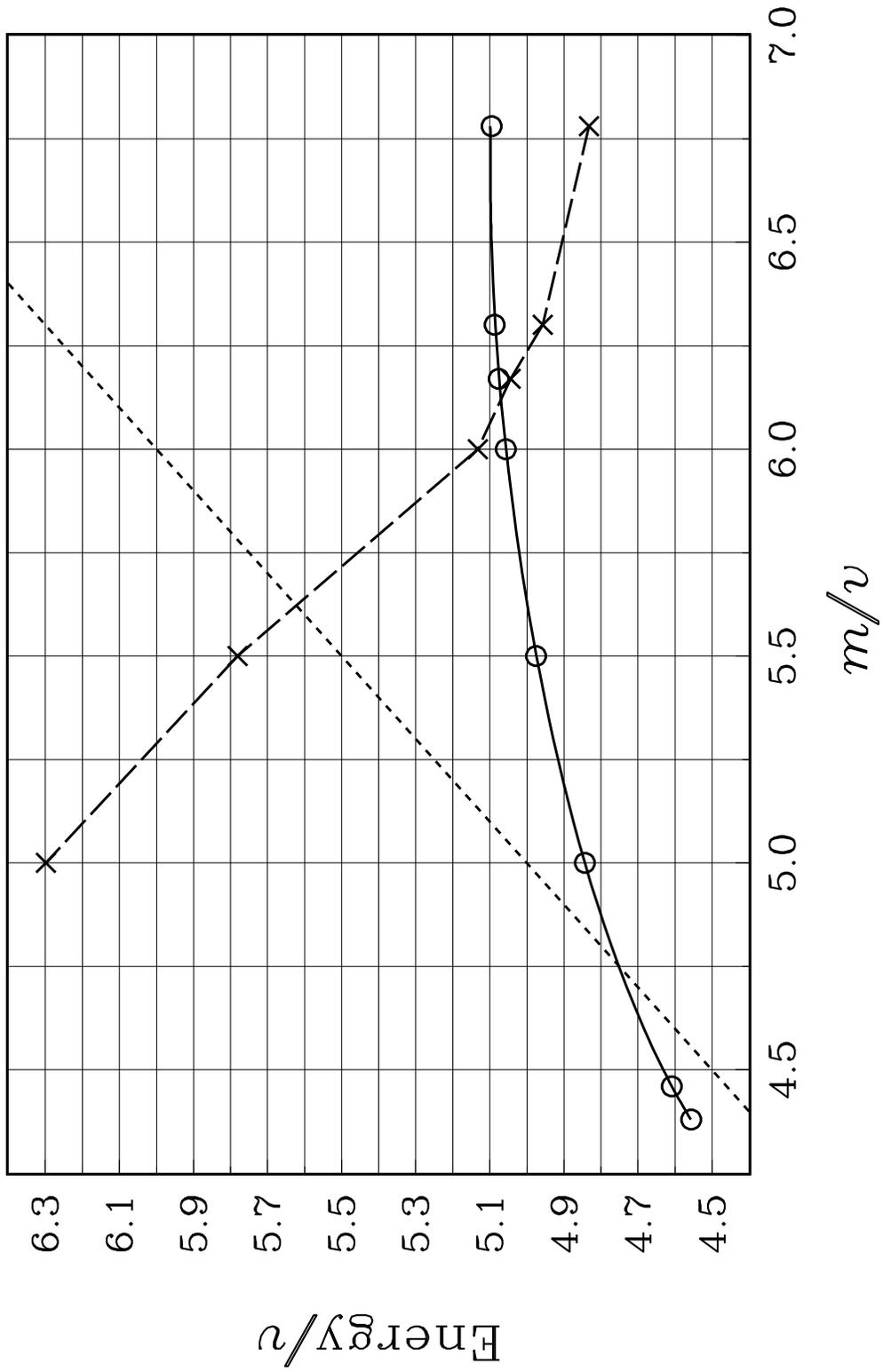
\item[Fig. 7] The minimal value of $E_{eff}$ (\ref{Eeff})
with $\omega_0=\Omega_0$ as a function of $m$ (solid line).
This curve consists of a few hundreds points obtained by searching for
$\Omega_0^{min}$ only. A few points found as the minima of $E_{eff}$
after generation of the effective valley are shown with the circles.
The plane wave energy $m$ is indicated with the short-dashed line.
The crosses connected with the long-dashed line show the value of
$E_{eff}$ at the left edge of the effective valley.
The numerical error seen in this value is due to
the poor statistics of the Monte Carlo procedure at small
$\Omega_0$.

\end{description}


\vspace{1cm}
\newpage

\begin{thebibliography}{}
\bibitem{Do} See, for example, A.
Dobado, Phys. Lett. {\bf B 237} (1990) 457.
\bibitem{col} G. 't Hooft, Phys. Rev. {\bf D 14} (1976) 3432;
 I. Affleck, Nucl. Phys. {\bf B 191} (1981) 429.
\bibitem{BK}J. Baacke and V. G. Kiselev,
Phys. Rev. {\bf D 48} (1993) 5648.
\bibitem{top}  F. Abe {\it  et al.}, FERMILAB-PUB-94/097E (1994).
\bibitem{bss} J. Baacke, H. So and A. Suerig, DO-TH-94/04 (1994).
\bibitem{Higgs}J. F. Guninon, H. E. Haber, G. Kane and S. Dawson,
The Higgs Hunter's Guide
(Addison-Wesley Publishing Company, Redwood City, CA, 1990)
\bibitem{triv} See ref. \cite{Higgs}. A more recent list of the
publications can be found, for example, in
B. A. Kniel, Phys. Rep. {\bf 240} (1994) 211.
We apologize to the authors of many interesting papers for such an
indirect citation.
\end{thebibliography}
\end{document}